\documentclass[]{emulateapj}
\newcommand{\HST}{\emph{HST}}
\slugcomment{version: 4/5/07}

\begin{document}

\title{Rest-Frame R-band Lightcurve of a $z \sim 1.3$ Supernova Obtained with Keck Laser Adaptive Optics
\label{chapter_sn}}

\author{J. Melbourne  \altaffilmark{1},
K.S. Dawson \altaffilmark{5},
D.C. Koo  \altaffilmark{1},
C. Max  \altaffilmark{1,2},
J. E. Larkin \altaffilmark{3},
S. A. Wright \altaffilmark{3},
E. Steinbring  \altaffilmark{4},
M. Barczys \altaffilmark{3}
G. Aldering \altaffilmark{5},
K. Barbary \altaffilmark{5,11},
M. Doi \altaffilmark{6}, 
V. Fadeyev \altaffilmark{5},
G. Goldhaber \altaffilmark{5,11},
T. Hattori \altaffilmark{7},
Y. Ihara \altaffilmark{6}, 
N. Kashikawa \altaffilmark{8},
K. Konishi \altaffilmark{9},
M. Kowalski \altaffilmark{5},
N. Kuznetsova \altaffilmark{5},
C. Lidman \altaffilmark{10},
T. Morokuma \altaffilmark{6}, 
S. Perlmutter \altaffilmark{5,11}, 
D. Rubin \altaffilmark{5,11},
D.J. Schlegel \altaffilmark{5},
A. L. Spadafora \altaffilmark{5},
N. Takanashi \altaffilmark{6},
N. Yasuda \altaffilmark{9}
}

\altaffiltext{1} {University of California Observatories/Lick Observatory, Department of Astronomy and Astrophysics, University of California at Santa Cruz, 1156 High Street, Santa Cruz, CA 95064. jmel, lotz, koo, max@ucolick.org}
\altaffiltext{2}{Institute of Geophysics and Planetary Physics, Lawrence Livermore National Laboratory, 7000 East Avenue, Livermore, CA 94550. max1@llnl.gov}
\altaffiltext{3} {Department of Physics and Astronomy, University of California, P.O. Box 951562, Los Angeles, CA 90095-1562. saw@astro.ucla.edu}
\altaffiltext{4} {Herzberg Institute of Astrophysics, National Research Council Canada, Victoria, BC V9E 2E7, Canada}
\altaffiltext{5}{E.O. Lawrence Berkeley National Laboratory, 1 Cyclotron Rd., Berkeley, CA 94720.}
\altaffiltext{6}{Institute of Astronomy, Graduate School of Science, University of Tokyo 2-21-1 Osawa, Mitaka, Tokyo 181-0015, Japan}
\altaffiltext{7} {Subaru Telescope, National Astronomical Observatory of Japan, 650 North A'ohaku Place, Hilo, HI 96720}
\altaffiltext{8}{Optical and Infrared Astronomy Division, National Astronomical Observatory of Japan, 2-21-1 Osawa, Mitaka, Tokyo, 181-8588, Japan}
\altaffiltext{9}{Institute for Cosmic Ray Research, University of Tokyo, 5-1-5, Kashiwanoha, Kashiwa, Chiba, 277-8582, Japan}
\altaffiltext{10}{European Southern Observatory, Alonso de Cordova 3107, Vitacura, Casilla 19001, Santiago 19, Chile}
\altaffiltext{11}{Department of Physics, University of California Berkeley, Berkeley, CA 94720}

\begin{abstract}
We present Keck diffraction limited $H$-band photometry of a $z \sim 1.3$ Type Ia supernova (SN) candidate, first identified in a Hubble Space Telescope (\HST) search
for SNe in massive high redshift galaxy clusters.
The adaptive optics (AO) data were obtained with the Laser Guide Star facility during four observing runs from September to November 2005.
In the analysis of data from the observing run nearest to maximum SN brightness, the SN was found to have a magnitude $H=23.9 \pm 0.14$ (Vega).
We present the $H$-band (approximately rest-frame $R$) light curve and provide a detailed analysis of the AO photometric uncertainties.  By constraining the aperture correction with a nearby ($4\arcsec$ separation) star we achieve  0.14 magnitude photometric precision, despite the spatially varying AO PSF.
\end{abstract}


\section{Introduction}
A remarkable outcome of the intermediate redshift ($z\sim0.5$) Type Ia supernova (SN) searches of
the 1990's was indication \citep{P98, G98, Schmidt98} and then the strong evidence \citep{Riess98, Perlmutter99} of a
cosmological acceleration \citep[for a review, see][]{Perlmutter03}.
In a decelerating universe, SNe should appear brighter in the past compared to their
expected brightness in a universe with zero acceleration.
What was discovered instead was that the SNe actually appeared fainter.
In the current cosmological model, this acceleration is driven by a negative-pressure component,
with density $\Omega_\Lambda$, commonly known as dark energy.
Analyses of larger data sets have confirmed this result \citep{ Tonry03, Knop03, Riess04}
and improved the constraints on cosmological parameters, indicating a dark-energy dominated universe
with $\Omega_M \sim 0.3$ and $\Omega_\Lambda \sim 0.7$.


The Nearby SN Factory \citep{SNfactory}, The Carnegie Supernova Project \citep{Hamuy06}, and the SDSS SN Survey \citep{SDSS} are engaged
in ground-based surveys of SNe at low redshifts, where calibration is most effective
using the full visible spectrum of relatively bright SNe.
Programs such as the SN Legacy Survey (SNLS) and ESSENCE have made significant progress in constraining
cosmological parameters with SNe in the intermediate redshift range $0.3 < z <0.8$ \citep{SNLS,essence}.
For SN searches at the highest redshifts, $z > 1$,
the high resolution and low sky background required to obtain high signal-to-noise measurements
are best achieved from a space-based telescope.
However, the current set of high-$z$ SNe is small ($< 20$ SNe at redshifts $z>1$).
Even with larger samples, the currently practical \HST\ observations of $z > 1$ SNe are expected to
be limited by the statistical and systematic uncertainty of extinction correction.

Using new data from the Hubble Space Telescope (\HST\footnote{Based on observations
made with the NASA/ESA Hubble Space Telescope, obtained from the data archive at the 
Space Telescope Institute. STScI is operated by the association of Universities for 
Research in Astronomy, Inc. under the NASA contract NAS 5-26555. 
The observations are associated with program 10496.}),
the Supernova Cosmology Project\footnote{http://supernova.lbl.gov/} is collaborating with members
of the IRAC Shallow Cluster Survey,
Red Cluster Survey\footnote{www.rcs2.org}, the Palomar Distant Cluster Survey\footnote{http://www-int.stsci.edu/$\sim$postman/hizclus/ftp.html},
and the XMM-Newton Distant Cluster Project \citep{XMM2}
to establish a new approach to measurements of high redshift SNe.
In a 219 orbit program, the Advanced Camera for Surveys (ACS) was used
to search for and observe SNe in recently discovered massive galaxy clusters \citep{RDCS,CL,WARPS1415,IRAC0223,XMM,RCS,Stanford,IRAC}
in the redshift range $0.9 < z < 1.5$.
These massive clusters provide not only a rich environment of potential high redshift SN hosts, but
also a well understood population of early type elliptical galaxies.
Observations of early-type galaxies have shown the dispersion in colors due to dust to be very small in
groups and clusters at low redshifts \citep{Hogg} and redshifts $z>1$ \citep{Blakeslee}.
This low dispersion has been attributed to a very low dust content in cluster ellipticals.
SNe discovered in cluster ellipticals are therefore expected to exhibit little or no dust extinction,
significantly reducing the largest statistical and systematic uncertainty in this redshift range.

With the advent of laser guide star (LGS) adaptive optics (AO)
on large ground based telescopes, there are now additional opportunities to
study high redshift SNe at rest-frame optical wavelengths.
AO generates high contrast imaging for point sources by forcing a significant percentage
of the light blurred by the atmosphere into the diffraction-limited core of the telescope point-spread-function (PSF).
For large telescopes, AO has the added benefit of high spatial resolution in the near-infrared,
a factor of four improvement over NICMOS when using Keck AO.
To compliment the \HST\ effort and to test the viability of future ground based follow-up of distant SNe,
we obtained Keck LGS AO images of one of the first SNe discovered in the \HST\ program to search for SNe in massive clusters.

In this paper we present AO photometry of a $z \sim 1.3$ Type Ia SN candidate. 
Section 2 outlines the SN discovery method and follow-up AO data.
Photometry, including detailed modeling of the AO PSF are presented in Section 3.
The $H$ band light curve from Keck LGS AO is shown in Section 4.
In section 5 we discuss the potential for  future AO studies of high redshift SNe.
The cosmological implications of this SN will be presented in a future paper upon completion of the analysis of the entire survey.

\section{Data}

The \HST\ search for SNe in high redshift galaxy clusters was performed primarily using the F850LP filter on ACS.
In  addition to the search orbits,
repeated \HST\ imaging covering the full SNe light-curves was obtained in the ACS F775W and F850LP filters.
Follow-up observations in the NICMOS F110W filters were scheduled when high redshift SNe were discovered.  

\begin{figure}[!t]
\begin{center}
\includegraphics[scale=0.7]{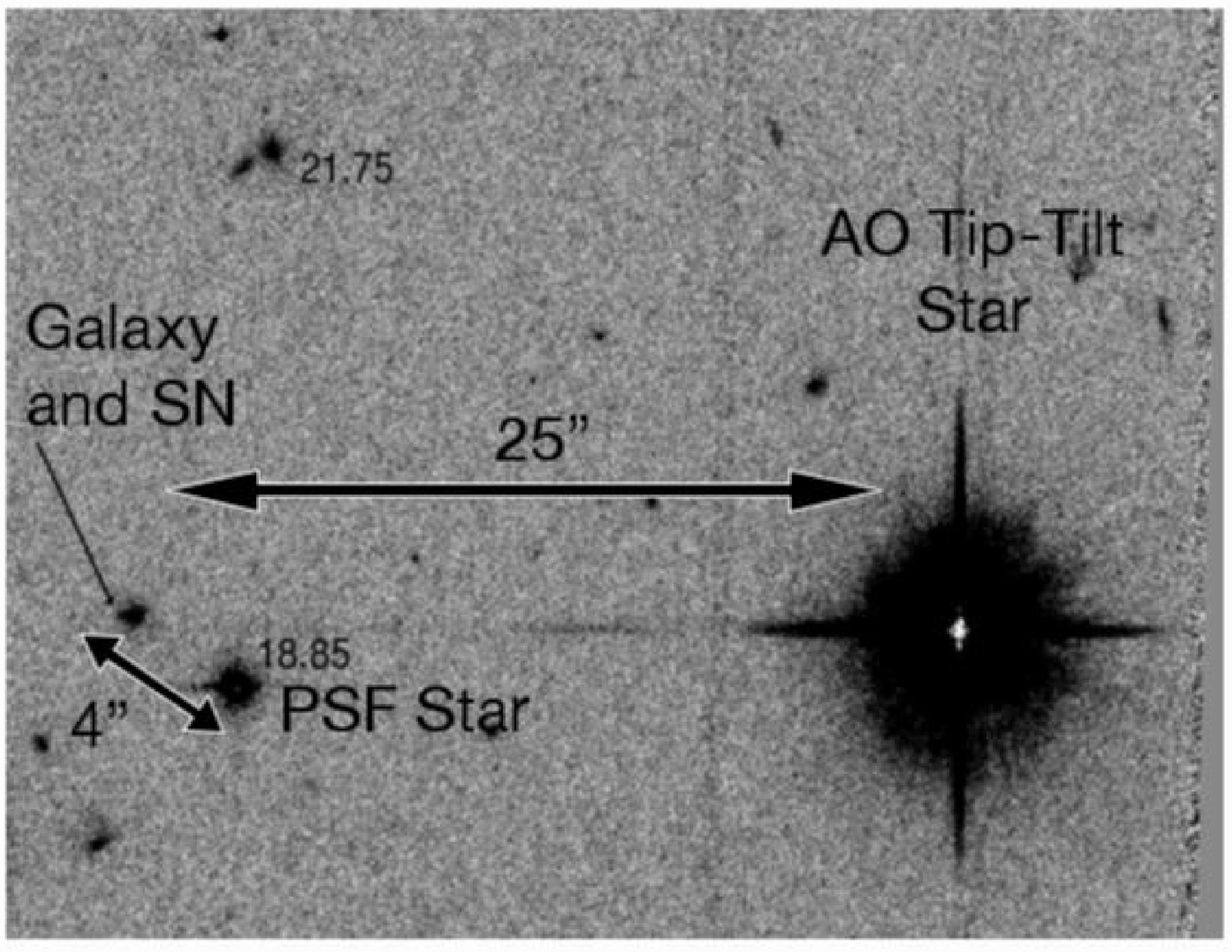}
\end{center}
\caption[Finder chart for SN O-006]{\label{fig:SN_finder} A finder chart for SN SCP05D6, nicknamed Maggie. This is an \HST\ ACS F850LP image obtained during the search
for SNe in massive galaxy clusters.
The SN was well positioned for laser-guide-star AO follow-up.  It was $25\arcsec$ from a 14th magnitude tip-tilt star and $4\arcsec$ from a 19th magnitude PSF star, ideal for tracking the time varying AO PSF and measuring aperture corrections. North is to the left. East is down.}
\end{figure}

The beginning of the SN search coincided with the Keck LGS AO observing runs of the Center for Adaptive Optics Treasury Survey
\citep[CATS,][]{Melbourneetal05}.  Because multiple high redshift SNe were discovered in the \HST\ search,
a target that was well suited for follow-up with adaptive optics could be selected.
For good AO performance, the Keck system requires the use of tip-tilt guide stars brighter than $R\sim18$ which
are used to correct the large scale tip-tilt of the wavefront.
Based on the frequency of these stars across the sky we anticipate the ability to observe $\sim20\%$ of
high galactic latitude targets.  As seen in an \HST\ image of the field, Figure \ref{fig:SN_finder},  SN SCP05D6, nicknamed Maggie, was only $25\arcsec$ from a $\sim14$th
magnitude tip-tilt guide star, well within the isokinetic patch of $60\arcsec$ radius \citep{Wizinowich06},
allowing for excellent AO correction.
In addition, this SN was only $4\arcsec$ from a  $\sim19$th  ($R$-band) magnitude star, ideal for constraining the complicated,
time-varying AO point-spread-function (PSF).  Based on the density of these fainter stars in the Great Observatories Origins Deep Survey \citep[GOODS,][]{Giavalisco04}, high galactic latitude targets should only have about a 2\% chance of being that close ($4\arcsec$) to a similarly bright star. 


Optical spectra of the SN and its host taken with FOCAS optical
low-resolution spectrograph \citep{Kash02} attached to
8.2-m Subaru telescope \citep{Iye04} show that the redshift of the SN host is
about 1.3.
The host is an elliptical galaxy populated with an old stellar population and
no indication of recent star formation.
The host properties indicate that it is highly unlikely that  SCP05D6 is
a core-collapse event and we conclude that it is a SN Ia.


\subsection{Follow-up Adaptive Optics Imaging}

Follow-up observations of Maggie were obtained at the W.M. Keck Observatory on four separate nights from September to November 2005. The SN was observed with Keck II using the NIRC2 instrument (K. Matthews et al. 2006, in preparation) in LGS mode. The laser was aligned with the galaxy. An $R=14.1$ star, $25\arcsec$ to the south, was used for tip-tilt correction. Observations were made in the $H$-band ($1.6\mu$m) with the narrow-field camera, with a pixel scale of $0.01 \arcsec$/pixel. Table \ref{table:obs_sum} details the dates, exposure times, and observing conditions of each run. The first three runs were under good seeing conditions ($<0.5 \arcsec$ in $K$), allowing good AO performance with a typical frame rate of $500\ Hz$ . During the final run, the seeing was more variable ($\sim0.5-1 \arcsec$ in $K$),
however, we were still able to run the AO system at $500\ Hz$. The October run suffered from intermittent cirrus cloud cover.  

Individual exposures were 60s, taken with a five position dither pattern ($1.5 \arcsec$ radius)
and two exposures at each position.
After each set of 10 exposures, an additional $\sim1\arcsec$ nod was applied to minimize the effects of bad pixels
and another ten exposures were taken following the same dither pattern.
For each exposure, we ensured that the SN, the galaxy, and the nearby star ($4\arcsec$ from the SN) remained within the $10\arcsec \times 10 \arcsec$ field of view of the camera.  Exposure times were long enough to allow alignment on the  star for post-process image combining. 

Immediately following the  SN observations, a United Kingdom Infrared Telescope (UKIRT) photometric standard star, FS6 \citep{Hawarden01},
was imaged at similar airmass (aimass of $\sim1.2$ for both standard and SN observations), setting the flux zeropoint for the observations. 

\begin{figure}[!t]
\begin{center}
\includegraphics[scale=0.7]{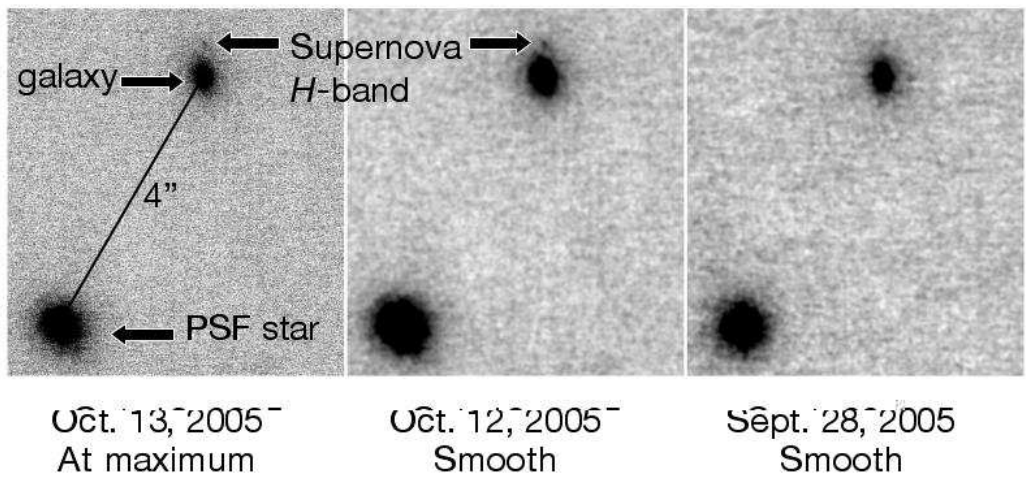}
\end{center}
\caption[Keck AO images of SN O-006]{\label{fig:SN_im}  Keck LGS AO images of Maggie.  The left image shows the SN near peak.  The middle is smoothed with a gaussian kernel of $0.06 \arcsec$ to enhance the contrast.  The right shows the SN two weeks before peak (also smoothed).  The right image has half the exposure time of the left image (30 min vs. 1 hour). North is up.  East is to the left. Note: the full AO images are $10\arcsec \times10\arcsec$, significantly larger than what is shown here. 
}
\end{figure}

Images were flatfielded with a twilight flat, then sky-subtracted using a scaled clipped mean sky image, obtained from the actual data frames.  Objects were masked out when creating the flat and sky images.  Finally, all frames from a given night were coadded by aligning on stars in the field, using a centroiding method to obtain image offsets.  Photometric zero points were obtained from observations of UKIRT standard star FS6 on each night.   The October run was not photometric.  Fortunately, the PSF star ($4\arcsec$ from the SN) was found to be non-time varying in the \HST\ data and so we tie the photometric zero point for the October night to the other 3 nights using the PSF star.  The September and October AO images are shown in Figure \ref{fig:SN_im}.  Note: the full images are $10\arcsec \times10\arcsec$ and only the region of interest is shown.

\section{Photometry}
Because of a spatially and time-varying point-spread-function (PSF), one of the great
challenges for AO science is to develop methods for precision photometry.
In this section, we demonstrate that photometric precision  of 0.14 mag at $H\sim 24$ is possible in an hour long Keck AO exposure.  In this case, the proximity (4" separation) of an $H=17.9$ magnitude star makes it possible to constrain the PSF of the SN and to determine an accurate aperture correction. We measure the photometric uncertainty by embedding model SNe into the galaxy halo. These model SNe were constructed from scaled versions of the nearby PSF star.  We also demonstrate that anisoplantism on these small spatial scales was negligible during the October observing run.

This section is subdivided into four subsections.  First we describe the galaxy subtraction and photometry of the PSF star.  Then we detail the aperture corrections from the PSF star and photometric uncertainties based on model SNe.  Next we discuss the issue of anisoplanetism, and finally we describe the photometry of SCP05D6.

\subsection{Galaxy Subtraction \label{sec:mod_gal} and Photometry of the PSF star}
\begin{figure}[!t]
\begin{center}
\includegraphics[scale=0.7]{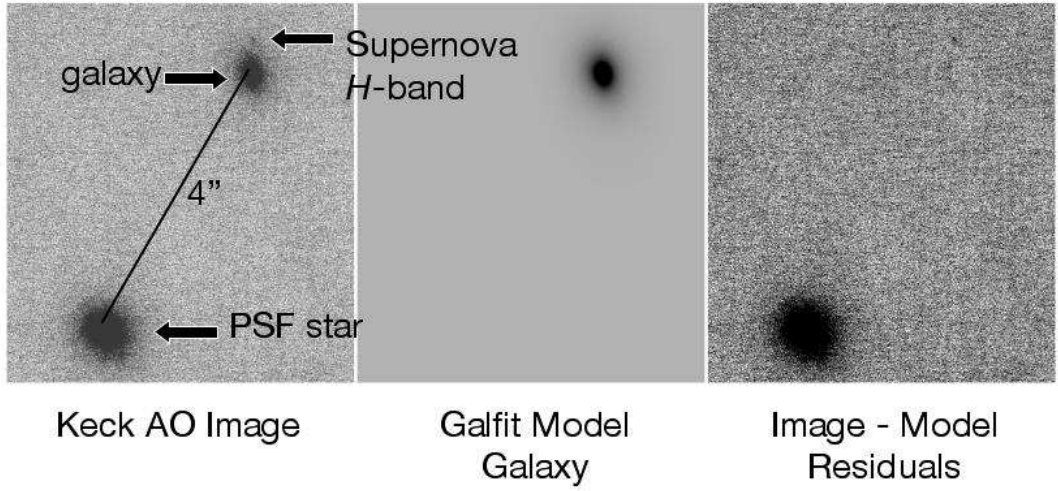}
\end{center}
\caption[Galfit model galaxy and residual image]{\label{fig:SN_model_im} Galfit \citep{Peng02} was used to fit for the galaxy light profile.
The model (middle image) is convolved with the input PSF.
The model galaxy is then subtracted from the input image (left image) to produce a residual image (right).
In this example, the galaxy has been removed, but the SN and PSF star remain.
}
\end{figure}

In order to accurately measure the magnitude of the SN, we must carefully account for the light of the host galaxy.
The fine pixel scale in the AO data ($0.01\arcsec$) provides excellent separation of the SN from the core of the galaxy
($\sim 40$ pixels, significantly better than \HST\ NICMOS $\sim 6$ pixels).
Using the GALFIT 2-d galaxy modeling routine \citep{Peng02} we search for the
best-fit galaxy Sersic model while simultaneously fitting the SN and PSF star.
The mean galaxy light profile, from all four epochs, is best fit by a Sersic index of $n=3.5$, consistent with an elliptical galaxy.
The H-band magnitude of the galaxy was found to be $19.22$.
Using GALFIT, the PSF convolved model galaxy is subtracted from the input image, leaving the SN and PSF star behind.
Figure \ref{fig:SN_model_im} shows the original science image, the galaxy model and the difference between the two.  

With the galaxy subtracted from the image, we measure the total magnitude of the nearby PSF star.  We use a curve of growth technique and measure the total magnitude at a radius of $1.8\arcsec$.  We find that the star has total $H$-band magnitude of $17.90 \pm 0.03$.  



\subsection{Aperture Corrections and Photometric Uncertainties \label{sec:apcor}}
At peak luminosity, the SN is roughly 1/100th the luminosity of the galaxy in the observed $H$-band. However, because it is a point source, it has high contrast against the host galaxy, 0.25 magnitudes higher surface brightness in an $0.1\arcsec$ radius aperture, $0.4\arcsec$ from the galaxy core.
Therefore, it should be measurable with small aperture photometry on the diffraction-limited core,
coupled with an accurate aperture correction from the nearby PSF star.
The accuracy and precision of this process will depend on several variables including,
1) the SN luminosity, 2) spatial variability in the galaxy luminosity profile (i.e. presence of dust or star forming regions within the galaxy),
3) exposure time, 4) the aperture size, and 5) spatial variation in the PSF.  

\begin{figure}
\begin{center}
\includegraphics[trim=0 0 0 150, scale=0.7]{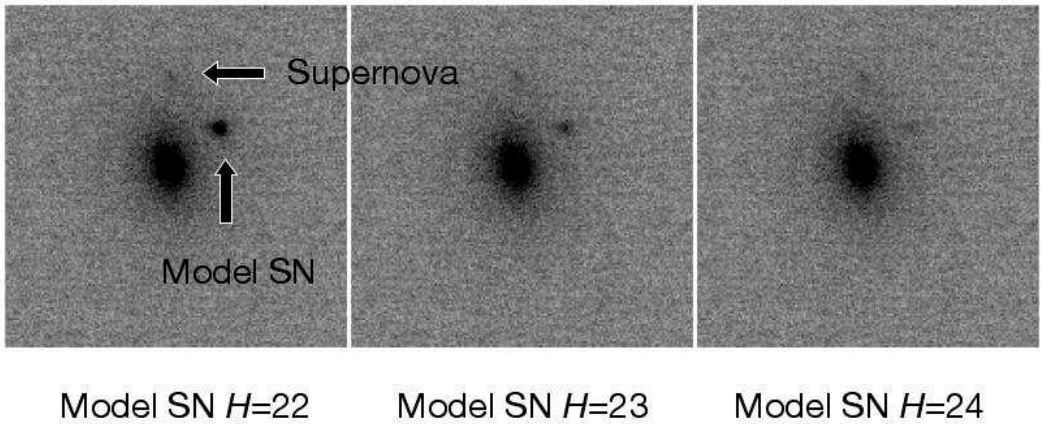}
\end{center}
\caption[Images of model SNe]{\label{fig:model_sn} Three model SNe with input magnitudes of $H=22$,  23, and 24 are shown.  They are placed at the same isophotal radius as the actual SN, but at a different PA.  The $H=24$ model is a close match to the luminosity of the actual SN in the October observation.  
}
\end{figure}

In order to address these issues, model SNe were imbedded into the galaxy halo.   Model SNe were generated by
scaling the PSF star to a range of $H=22 - 26$ mags, in increments of 0.25 mags.
Model SNe were placed in the galaxy at the same isophotal radius as SCP05D6, at twenty-one different position angles (PA)
relative to the host galaxy.
Figure \ref{fig:model_sn} shows three of these models imbedded in the October image.  We chose to do the modeling on the October image, because it was near the peak of the SN luminosity.

\begin{figure}[!t]
\begin{center}
\includegraphics[trim=0 30 0 40,scale=0.7]{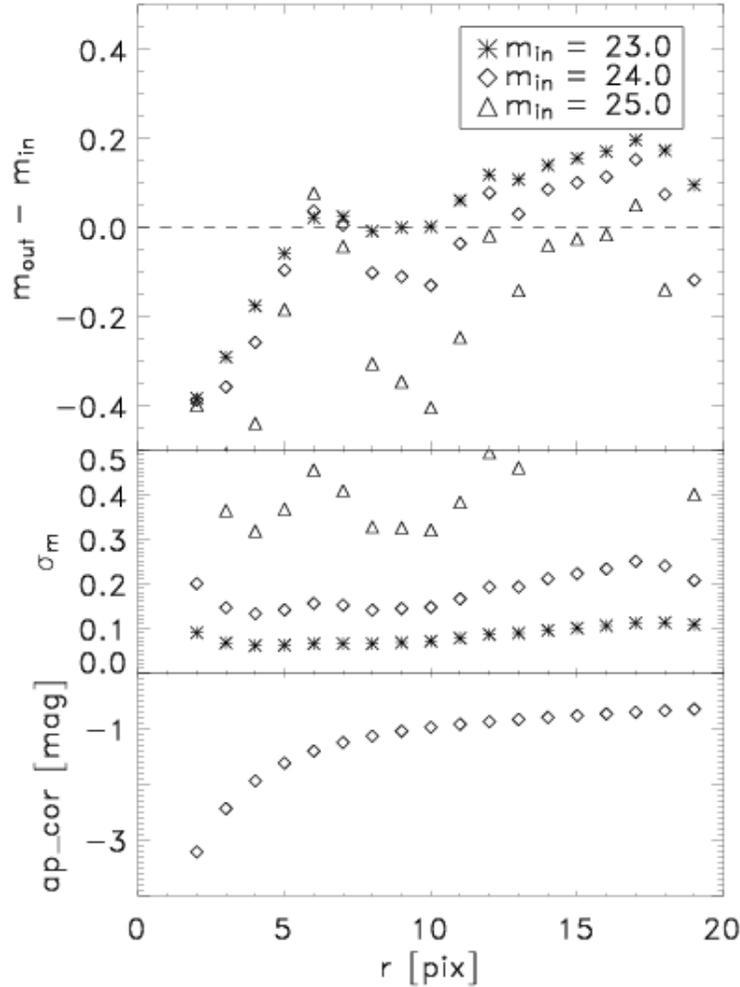}
\end{center}
\caption[Aperture photometry of three example model SNe]{\label{fig:model_ap_phot}  Aperture corrected photometry of three sample model SNe.  The top panel gives the difference between  measured and input magnitude as a function of aperture radius.  The middle panel shows the photometric uncertainty as a function of aperture radius.  The bottom panel shows the aperture correction as measured from the nearby PSF star.  For SN models with $H < 24.25$, the most accurate photometry is from apertures with $ r\sim10 $ pixels ($0.1 \arcsec$).  This is roughly twice the FWHM of the AO PSF.  For fainter sources, analysis of the full set of models indicates that the most accurate photometry is from smaller radius apertures, $r\sim6$ pixels ($0.06 \arcsec$), or roughly the FWHM of the PSF.
}
\end{figure}

After adding a model SN to an image, we subtracted the galaxy following the method outlined
in Section \ref{sec:mod_gal}.
We then performed aperture photometry on the model SNe and generated aperture corrections from the nearby star. Sky estimates were made from additional apertures placed at different PAs around the galaxy at the same isophotal radius as the SN and model.  
Figure \ref{fig:model_ap_phot} shows how the accuracy of the aperture corrected photometry (top)
and photometric uncertainty (middle) depend on aperture size for three sample models.
Photometric uncertainty was calculated from the Poisson statistics of the measured flux of the SN, host galaxy, and sky within the aperture.  The readout noise (9.5 e$^-$) was added in quadrature.
The Poisson sky contribution was estimated from the standard deviation of pixels in the sky apertures. The aperture correction used for each aperture size is given in the bottom panel of figure 
\ref{fig:model_ap_phot}.

For small ($r < 0.05 \arcsec$) and large ($r > 0.15 \arcsec$) radii the aperture photometry is not accurate. It is difficult to get an accurate aperture correction for small apertures.
Large apertures suffer from imperfections in galaxy subtraction.  For brighter models, $H < 24.25$, the most accurate photometry comes from apertures with radii between $0.09\arcsec$ and $0.11 \arcsec$, roughly twice  the size of the full-width-half-maximum (FWHM) for the AO PSF.  Analysis of the full set of fainter models, $H\geq 24.25$, indicates that smaller apertures are more accurate for these stars because they contain less sky.  For faint systems we use a $0.06\arcsec$ aperture.

\begin{figure}[!t]
\begin{center}
\includegraphics[trim=0 30 0 40,scale=0.7]{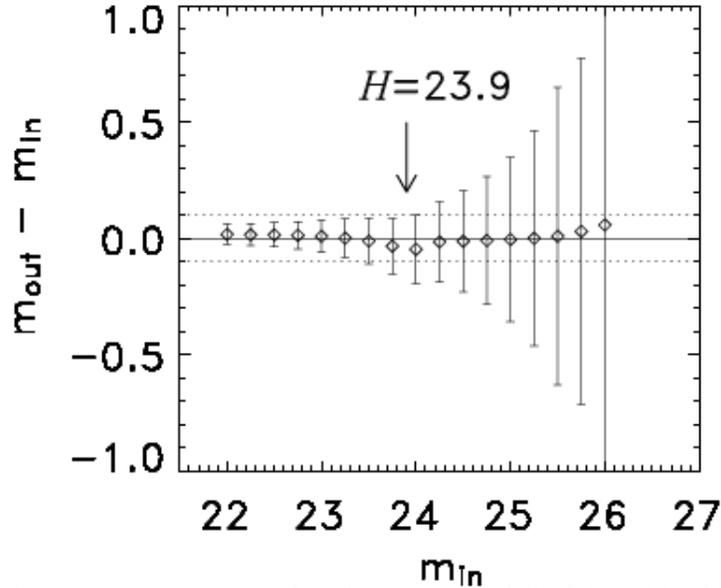}
\end{center}
\caption[Mean photometry of 420 model SNe]{\label{fig:model_tot_phot} The mean difference between measured magnitude and input magnitude for  SNe models, plotted against input magnitude.  For SN models with $H < 24.25$, the photometry is from apertures with $ r=10 $ pixels ($0.1 \arcsec$).  For fainter sources, the photometry is from smaller radius apertures, $r=6$ pixels ($0.06 \arcsec$). Twenty-one  models were produced for each input magnitude.  The error-bars are the standard deviation of the photometry of those 21 models.  The photometry is accurate to within 14\% for $H=24$.  An arrow indicates the location of the measured at peak $H$-band photometry of Maggie.
} 
\end{figure}

Figure \ref{fig:model_tot_phot} summarizes the photometry of the 360 model SNe.  For each input magnitude, we plot the mean difference between input and measured magnitude of the models.  For $H < 24.25$, output magnitudes are measured from $0.1\arcsec$ (10 pixel) radius apertures. For fainter systems, we use $0.06 \arcsec$ apertures. The error-bars are the standard deviation of the difference between model input and output magnitudes.  The photometry is measured to within $14\%$ for $H\sim24$ (Vega) in an hour long science exposure.  These variations are on the order of the formal photometric uncertainties indicating that systematic variations in the galaxy light or sky background are small.  At $H=24$ there is an offset between our input model and measured photometry of $\sim0.04$ mags.  As this is less than half the photometric uncertainty, we do not perform any additional correction to the final magnitudes. 

\subsection{Anisoplanatism}

When making the model SNe, we neglected to account for anisoplanatism.
It was assumed that the PSF did not change over the $4\arcsec$ separation between the SN and the nearby star.
In order to test if this is a valid assumption, we employed a simple anisoplanatic
AO PSF model to investigate its effect on the aperture corrections.
Artificial PSFs were generated with an LGS AO simulator program written by E. Steinbring \citep[see][]{Steinbring05}.
The model is based on a single turbulent layer at a fixed atmospheric height.  The inputs to this program include the relative position of the laser spot and the tip-tilt guide star, the uncorrected $V$-band seeing, the central wavelength of the observation, the mean height of the turbulence, the size of the telescope, the number of AO actuators and the system frame rate. The output is an artificial star field with a spatially varying PSF.  

The program input parameters closely match the conditions during observations at Keck.
The program output is the image of an artificial star field with stars separated in a grid by $4\arcsec$.
The aperture correction analysis outlined in Section \ref{sec:apcor} was applied to the artificial star field.
The anisoplanatism is primarily set by the mean height of turbulence and $r_0$ (or seeing).
If turbulence is restricted to below 6 km, the typical difference in aperture corrections for stars separated by $4\arcsec$ is 0.03 magnitudes
using apertures with radii as small as $0.1 \arcsec$.
This variation is significantly smaller than the expected photometric uncertainty for our faint SN image.  However if the seeing is poor and the mean height of the turbulence is 6 km or higher, the variations in aperture correction over a $4 \arcsec$ separation can reach 0.15 magnitudes and greater.  

Fortunately, we have two types of observations that characterize the atmospheric conditions during our run.    First, observations of a star field obtained with the NIRC2 wide camera in the $K'$ filter an hour after completing our SN observations.  The star field was observed as part of a separate extra-galactic AO program, and was not specifically designed for the SN program.  Despite this, we can use it to learn something about the atmospheric conditions on the night of Oct 13.  Aperture corrections, calculated in the same manner as Section \ref{sec:apcor}, from stars separated by up to $10\arcsec$, are constant to within 0.05 magnitudes.  This indicates good seeing and a mean turbulence height likely below 6 km. In which case, the aperture corrections for our SN photometry are reasonable.

 The second set of data comes from the Thirty Meter Telescope (TMT) site testing team.  M. Schoeck et al. were using a Multi-Aperture Scintillation Sensor \citep[MASS][]{MASS98} to measure the turbulence profile on Mauna Kea during three of our observing runs.  They provided us with a measure of the high altitude seeing (excluding the ground layer) for the times of our SN observations.  Mean seeing from MASS is recorded in Table 1.  For the October run, the MASS results indicate that that the seeing was very good, and was stable throughout the SN and star field observations, varying between 0.3 and 0.7 arcsec.  Because the high altitude turbulence was light we again expect anisoplantism was small on small spatial scales.

\subsection{SN Photometry}
\begin{figure}
\begin{center}
\includegraphics[scale=0.6]{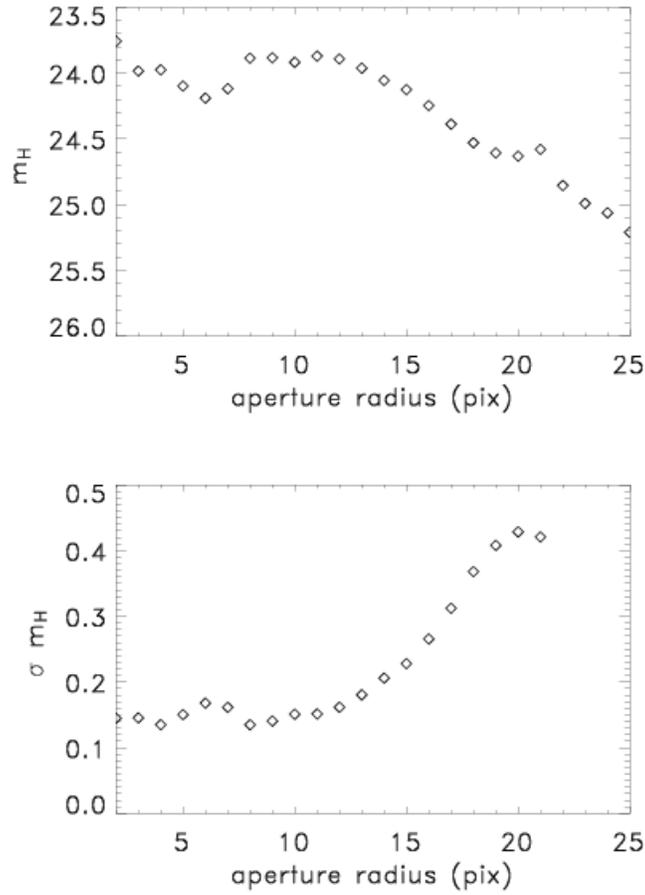}
\end{center}
\caption[Photometry of SN O-006]{\label{fig:sn_phot} $H$-band photometry for Maggie on the UT date, October, 13 2005.  The measured magnitude reaches a plateau for  apertures with $r\sim 10$ pixels, as was expected from the model SN.  These radii also correspond to the minimum in the photometric uncertainty.  As with the model SNe, we used the 10 pixel radius aperture for our final photometry giving a total magnitude of $H= 23.9\pm 0.14$, where the error-bars come from the model simulations described in the previous section. 
}  
\end{figure}

Figure \ref{fig:sn_phot} shows aperture photometry of Maggie from the October run, after aperture corrections have been applied.  As with the model SNe, the photometry of Maggie levels off with aperture radii of 9-11 pixels, meaning that at these radii the aperture corrections are producing consistent results.  Taking the magnitude from the $r=0.1\arcsec$ (10 pixels) aperture, Maggie has an H-band magnitude of $23.9\pm 0.14$, where  the uncertainty is estimated from the model simulations described in Section \ref{sec:apcor}.   

The measurements from each observing run can be found in Table \ref{table:obs_sum}.
The observed $H$-band light curve drops by about half a magnitude in the September and November observations
from the peak luminosity in October, resulting in photometric uncertainties in excess of 0.2 magnitudes. Analysis of the model SNe indicates that smaller apertures improve the accuracy for faint SNe.
Using apertures with a radius of $0.06 \arcsec$ we estimate the photometry of the September and November 25th observing runs,
$H=24.4 \pm 0.22$, and $H=24.8 \pm 0.28 $ respectively.
Note, the November 12th run was the lowest signal-to-noise observation and we were unable to get an estimate of the SN luminosity.  

\section{Results}
\begin{figure}
\begin{center}
\includegraphics[scale=0.6]{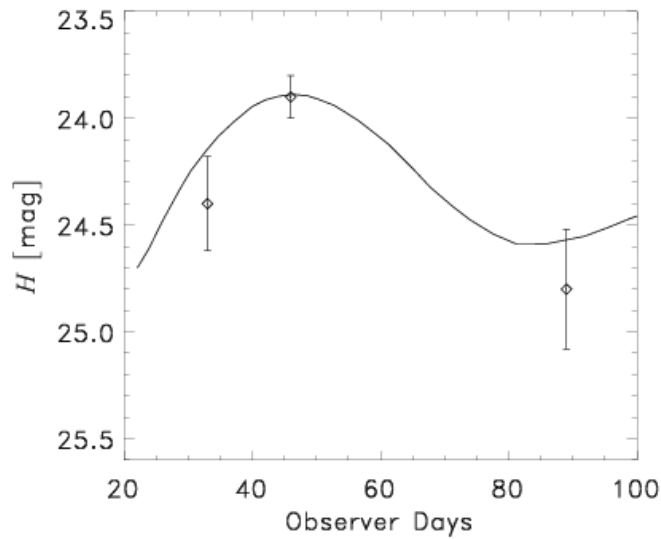}
\end{center}
\caption{\label{fig:light_curve} The supernova lightcurve of Maggie, in the observed  $H$-band.  An example lightcurve based on SN 1995D ($z=0.0066$) is also shown. The light curve was K-corrected to the corresponding redshift and broad band filters using a time series of averaged Type Ia SN spectra \citep{nugent}.  From the \HST\ observations,  rough estimates were made for reddening, time of maximal light, peak luminosity and stretch.
}
\end{figure}

Figure \ref{fig:light_curve} shows the $H$-band (rest frame $R$) light curve for Maggie obtained with Keck AO imaging.
An example SN template based on the light-curve of SN 1995D ($z=0.0066$)  is scaled and over-plotted.
The template light curve was K-corrected to the corresponding redshift and broad band filter \citep{nugent}.
Based on the \HST\ ACS F850LP and NICMOS F110W photometry
a rough estimate was made for reddening, time of maximal light
peak luminosity and stretch (stretch represents a linear transformation of the time axis: $t^\prime = t\cdot s$).
Because of the large uncertainties on the two fainter $H$-band data points,
we do not provide a formal fit to the AO observations.
Preliminary results indicate that the $H$-band photometry near max matches, to within the uncertainties, the expectations from the \HST\ $J$-band observations.  This suggests that the host galaxy does not contain large amounts of dust, which is again consistent with the elliptical morphology. Light curves from \HST\ and the Hubble diagram for Maggie will be published in a forthcoming paper from the Supernova Cosmology Project. 


\section{Discussion and Conclussions}
Using Keck LGS AO, we obtained follow-up observations of a high-$z$ SN discovered in an \HST\ search for SNe in massive galaxy clusters. During a one-hour observation, we obtained 14\% photometry of $H\sim24$ supernova near peak luminosity embedded in a distant early-type galaxy.
When combined with the \HST\ observations, the AO data will help constrain the dust content
of the galaxy and SN, important for improving the accuracy of the measured light curves.  

In this example we were able to well characterize the AO PSF.  In addition the host galaxy was a relatively dust free elliptical.  Therefore our uncertainties were primarily set by the photometric error, i.e. the faintness of the source, the high IR background, and the sensitivity of our detector.  As a result, we could have improved our photometry simply by increasing our exposure time.  A four hour exposure would have increased the signal-to-noise ratio by a factor of 2, giving a photometric uncertainty of $\sim0.07$ mags for the at peak observation.  At this level, the photometric uncertainty is on the order of the aperture correction uncertainties set by the time varying PSF, which in our case were small, $\lesssim 0.05$ mags.  Similar increases in signal-to-noise ratio will be possible with the next generation of large telescopes ($\sim30m$ diameter).  However, in order to reach the level of precision now possible from space, 1\%-3\% photometry, we will need to improve on our AO PSF characterization methods.   

We want to stress that adaptive optics was necessary to make this observation from the ground. For instance to reach a S/N ratio of 7 with ISAAC (the seeing limited IR imager on the VLT) for an object with $H=24$  one needs 20 hours, and that is without a host. The presence of a bright host means that a reference image of the host without the SN is mandatory. If this image is as deep as the one with the SN, then total integration time doubles to 40 hours and the S/N ratio of the SN in the host galaxy subtracted image drops to 5.

Accurate photometry with AO requires knowledge of the real-time PSF. In the case of this particular SN, a nearby faint star was ideal for tracking PSF variations and computing accurate aperture corrections. This SN was specifically chosen for AO follow-up because of this arrangement.
We caution that most observations will not have a nearby PSF star.  Only 2\% of high galactic latitude targets will be $4\arcsec$ from a similarly bright star.
It is therefore important to develop additional techniques for tracking the real-time, spatially-varying PSF.  Groups such as the Center for Adaptive Optics Treasury Survey \citep[CATS][]{Melbourneetal05} are developing these techniques now.
For example, \cite{Steinbring05} explore a method using observations of a star cluster to constrain
the spatial variation of the PSF. Britton et al. (2006) show how this can be done with auxiliary data from a Differential Image Motion Monitor (DIMM). These methods are promising and could pave the way for precision AO photometry of distant SNe.       

AO follow-up of SNe is also limited by the availability of tip-tilt guide stars.
For good performance with the current Keck AO system, tip-tilt stars must be brighter than $R=18$.
This constraint excludes follow-up of roughly 80\% of high galactic latitude SNe.
So, while AO improves on the resolution capabilities of \HST\ NICMOS,
NICMOS remains necessary for the rest-frame optical follow-up of most high-$z$ SNe.
This may change with the next generation Keck AO system which is being designed
with near-infrared tip-tilt wavefront sensors.  In this new system, tip-tilt stars will be AO corrected, resulting in high contrast above the background.  Therefore the new system will be able to function with fainter tip-tilt stars, opening more night sky to AO observations.

An additional constraint on current AO systems is the small field of view of the cameras.
The sky area corrected in a single exposure is smaller than one square arc-minute.
The small field of view makes AO searches for distant SN impractical.
Future systems with multiple correcting elements, such as multi-conjugate adaptive optics (MCAO),
may be able to increase the corrected area of a single exposure to roughly 4 square arc-minutes.
With a larger detecting area, and a search focused on high-redshift galaxy clusters,
future AO systems may prove useful in the discovery and follow-up of distant SNe.

\acknowledgments

This work has been supported in part by the NSF Science and 
Technology Center for Adaptive Optics, managed by the University of California (UC) at
Santa Cruz under the cooperative agreement No. AST-9876783.

Financial support for this work was also provided by NASA through program GO-10496 from the Space Telescope Science Institute, which is operated by AURA, Inc., under NASA contract NAS 5-26555. This work was supported in part by the Director, Office of Science, Office of High Energy and Nuclear Physics, of the
U.S. Department of Energy under Contract No. AC02-05CH11231. This work is supported in part with a scientific research grant 
(15204012)by the Ministry of Education, Science, Culture, Sports, 
Science and Technology
of Japan.

Subaru observations were collected at Subaru Telescope, which is 
operated by the National Astronomical Observatory of Japan.

The laser guide star adaptive optics system was funded by the W. M. Keck
Foundation. The artificial laser guide star system was developed and
integrated in a partnership between the Lawrence Livermore National Labs
(LLNL) and the W. M. Keck Observatory. The laser was integrated at Keck with
the help of Curtis Brown and Pamela Danforth. The NIRC2
near-infrared camera was developed by CalTech, UCLA and Keck
(P.I. Keith Matthews). The data presented herein
were obtained at the Keck Observatory, which is operated as a
scientific partnership among the CalTech, UC and NASA.  
This work is supported in part under the auspices of the 
US Department of Energy, National Nuclear Security Administration and by 
the LLNL under contract W-7405-Eng-48.

The authors wish to recognize and acknowledge the very significant cultural
role and reverence that the summit of Mauna Kea has always had within the
indigenous Hawaiian community.  We are most fortunate to have the
opportunity to conduct observations from this superb mountain.


\clearpage

\clearpage

\begin{deluxetable}{ccccccc}
\tablewidth{0pt}
\tabletypesize{\small}
\tablecaption{\label{table:obs_sum} The Keck AO Observing Summary for SN SCP05D6.  }

\tablehead{\colhead{UT Date} & \colhead{Exp. Time} & \colhead{Sky} & \colhead{Mean Seeing \tablenotemark{a} }&  \colhead{PSF FWHM \tablenotemark{b}} &\colhead{Photometry} & \colhead{SN $H$-band magnitude} \\
\colhead{} & \colhead{(s)} & \colhead{Conditions} &\colhead{[$\arcsec$]}&  \colhead{($\arcsec$)} &\colhead{Aperture Size $[\arcsec]$} & \colhead{(Vega)}}
\startdata
Sept. 29, 2005 & 1800 & photometric & -    & 0.052& 0.06    & $24.4 \pm 0.22$ \\
Oct. 13, 2005 & 3600 & some cirrus & 0.40  & 0.053& 0.10    & $23.9 \pm 0.14$ \\
Nov. 12, 2005 & 1740 & photometric& 0.45   & 0.053& -       & - \\
Nov. 25, 2005 & 3300 & photometric& 0.35   & 0.064& 0.06    & $24.8 \pm0.28$ \\
\enddata
\tablenotetext{a}{Measured from MASS by Matthias Schoeck and the TMT site testing team.  Note: MASS does not measure ground layer turbulence, so this represents the high altitude seeing.} 
\tablenotetext{b}{Measured from a Gaussian fit.}

\end{deluxetable}

\end{document}